\documentclass[aps,pra,twocolumn,showpacs,floatfix]{revtex4}
\usepackage{amsmath,graphicx}
\newcommand{\ket}[1]{\vert #1 \rangle}
\newcommand{\bra}[1]{\langle #1 \vert}

\newcommand{\HH}{{\hbox{\small HH}}}

\newcommand{\VV}{{\hbox{\small VV}}}
\newcommand{\PP}{{\hbox{\small\sc pur}}}
\newcommand{\D}{{\hbox{\small\sc d}}}

\begin{document}
\title{
Nonlocal compensation of pure phase objects with entangled
photons}
\author{Simone Cialdi}\email{simone.cialdi@mi.infn.it}
\affiliation{Dipartimento di Fisica dell'Universit\`a degli Studi
di Milano, I-20133 Milano, Italia.}
\affiliation{INFN, Sezione di Milano, I-20133 Milano, Italia.}
\author{Davide Brivio} \email{davide.brivio@unimi.it}
\affiliation{Dipartimento di Fisica dell'Universit\`a degli Studi
di Milano, I-20133 Milano, Italia.}
\author{Enrico Tesio}\email{enrico.tesio@strath.ac.uk}
\affiliation{SUPA, Department of Physics, University of Strathclyde, Glasgow G40NG,
Scotland, UK}
\affiliation{Dipartimento di Fisica dell'Universit\`a degli Studi
di Milano, I-20133 Milano, Italia.}
\author{Matteo G. A.~Paris}\email{matteo.paris@fisica.unimi.it}
\affiliation{Dipartimento di Fisica dell'Universit\`a degli Studi
di Milano, I-20133 Milano, Italy}
\affiliation{CNISM, Udr Milano, I-20133 Milano, Italy}
\date{\today}
\begin{abstract}
We suggest and demonstrate a scheme for coherent nonlocal compensation
of pure phase objects based on two-photon polarization and momentum
entangled states. The insertion of a single phase object on one of the
beams reduces the purity of the state and the amount of shared
entanglement, whereas the original entanglement can be retrieved by
adding a suitable phase object on the other beam.  In our setup
polarization and momentum entangled states are generated by spontaneous
parametric downconversion and then purified using a programmable spatial
light modulator, which may be also used to impose arbitrary space
dependent phase functions to the beams. As a possible application, we
suggest and demonstrate  a quantum key distribution protocol based on
nonlocal phase compensation.
\end{abstract}
\maketitle
\section{Introduction}
In what is usually referred to as {\em ghost imaging} the coherent
imaging of an object is achieved with incoherent light upon exploiting
the spatial correlations between two light beams \cite{aga08}. The
object interacts with one of the beams and an image of the object is
built up by scanning the other beam. Ghost imaging may be obtained
either with classically correlated beams \cite{ben02,fer05,kat09}, as
those obtained by splitting the light from a (pseudo) thermal source, or
with entangled beams \cite{rib94,pit95,abo01}, as those obtained by
parametric downconversion. In the latter case one may achieve in
principle higher visibility.
\par
For objects which modify only the amplitude of light an image may be
obtained with a single spatially incoherent beam upon measuring the
autocorrelation function in the far field, without the need of ghost
imaging. This is no longer possible when the object is also modifying
the phase of the beam. In particular, it is of interest to investigate
ghost imaging in the extreme case of pure phase objects, i.e. objects
altering only the phase information carried by the beam.  Phase objects
are also of intrinsic interest in quantum information processing, since 
they introduce reversible unitary operations. 
\par
Ghost imaging of pure phase objects has been extensively analyzed
theoretically and experimentally using both classically or quantum
correlated beams
\cite{ben02,fer05,kat09,rib94,pit95,abo01,Zha08,Gan11,aga03,serg04}. 
Related effects connected with (nonlocal) dispersion 
cancellation have been investigated as well, both in the temporal and
the spatial domains \cite{Ste92,Bon08,Min09,Was10,Fra92,Bae09}.
In this paper we suggest and demonstrate experimentally a scheme 
to achieve coherent nonlocal compensation/superposition of pure phase 
objects, also paving the way for the reconstruction of the overall 
phase function imposed 
to the two beams. Our scheme 
is based on two-photon polarization and momentum entangled states, 
which are generated by spontaneous parametric downconversion and
purified using a spatial light modulator. The same device is also used 
to introduce arbitrary phase functions on the two beams, which represent
arbitrary phase objects.  In our setup the insertion of a single
phase object on one of the beams reduces the purity of the state and the
amount of shared entanglement, whereas the original entanglement can be
retrieved by adding a suitable phase object on the other beam. The
image of both single or double phase objects can be thus obtained by
scanning the coincidence counts on one of the two beams.
As a possible application, we also suggest and demonstrate a protocol for 
quantum key distribution based on nonlocal phase compensation.
\par
The paper is structured as follows. In the next Section we describe 
in details our experimental setup and the properties of the  
two-photon entangled states that are generated. In Section \ref{s:dpi} 
we analyze in some details phase imaging and nonlocal phase compensation of 
phase objects. Finally, in Section \ref{s:qkd}, we suggest a quantum key 
distribution protocol based on phase compensation and report about 
its experimental implementation. Section 
\ref{s:out} closes the paper with some concluding remarks.  
\section{Experimental setup}\label{s:exp}
\begin{figure}[h!]
\includegraphics[width=0.47\columnwidth,angle=-90]{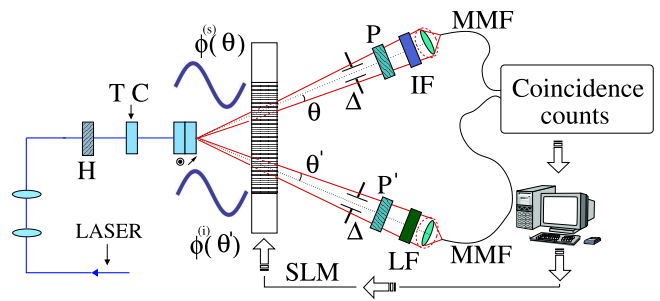}
\caption{(Color online) Schematic diagram of experimental setup. A
linearly polarized cw laser diode at $405\,$nm pumps a couple of BBO
crystals cut for Type-I downconversion. 
The horizontal and vertical photon pairs are balanced by a half
wave-plate set along the pump path, whereas an additional BBO crystal
(TC) is placed on the pump path to compensate the temporal delay. Signal and idler
beams travel through the SLM, which provides entanglement
purification and imposes the space dependent phase functions
$\phi^{(s)} (\theta), \phi^{(i)} (\theta^\prime)$, and then 
are spatially selected by two irises
and two slits set at $D=500\,$mm with $\Delta x=5\,$mm
($\Delta=10\,$mrad). An interference filter IF (FWHM $10\,$nm) set on
the signal path selects the spectral width, while a long-pass filter LF,
cut-on wavelength $715\,$nm, is set on the idler path in order to reduce
the background. Photons are focused in two multi-mode fibers (MMF) and
sent to single-photon counting modules. Polarizers at the angles
$\pi/4$ and $3\pi/4$ or $\pi/4$ and $\pi/4$ are inserted to
measure visibility whereas a quarter-wave plate, a half-wave plate and a
polarizer (not shown in the figure) are used for the tomographic
reconstruction.}
\label{exp_setup} 
\end{figure}

In our setup (see Fig. \ref{exp_setup}) 
a two-qubit polarization entangled state is produced by
type-I downconversion from a couple of crystals (beta-barium
borate, BBO) in a non-collinear configuration
\cite{har92,Kwiat99,gen05}. Pairs of correlated photons are generated, 
distributed on a broad angular and spectral ranges, which are 
determined by the crystal length~\cite{cia08}. Upon expanding the
transverse momentum conservation condition to the first order, it can be
shown that the angular and the spectral degrees of freedom are connected
by the relation 
\begin{equation}
\label{eq:theta-omega}
\theta'=-\theta+\gamma\omega\;,
\end{equation} 
where $\theta,\theta'$ are the signal and idler shifts from the central
emission angles ($\Theta_0 \simeq \Theta_0'\simeq 3^\circ$ in our case), 
$\omega$ is the signal shift from the central frequency of the downconverted
beams, and $\gamma$ is a constant depending only on the signal
central frequency and angle ~\cite{cia10APL,cia10}. 
\par
The state at the output of the crystals may be written as 
\begin{align}
\ket{\psi}\!\propto\! \int\!\!\!\int\!\! d\theta d\theta'
&g(\theta,\theta')\notag\times\\
&\left[ \ket{H\theta}\ket{H\theta'}+e^{\imath \Phi(\theta,\theta')}
\ket{V\theta}\ket{V\theta'}\right]
\label{eq:state}
\end{align}
where the overall angular distribution 
$$g(\theta,\theta')=f(\theta,\theta')\,
T[\omega(\theta,\theta')]\,,
$$
contains the angular distribution $f(\theta,\theta')$ 
due to the phase-matching conditions
and the transmissivity 
$T[\omega]\equiv T[\omega(\theta,\theta')]$ 
of an interference filter set on the signal arm.
The ket $\ket{P\theta}$ denotes a single photon state emitted with
polarization $P=H,V$ at angle $\theta$ ($\theta'$) along the signal
(idler) arm, and 
the integrations ranges from $-\frac12 \Delta$ to $\frac12 \Delta$, 
$\Delta$ being the angular aperture of two slits placed along the 
paths of the downconverted beams (see Fig.~\ref{exp_setup}). 
\par
Experimentally, 
some care must be taken in order to spatially superimpose the 
$\ket{\HH}$ and $\ket{\VV}$ downconversion beams, so
that the angular distribution $f(\theta,\theta')$ is actually the same
for the two components of the entangled state. From
Eq.~(\ref{eq:theta-omega}) one sees that for a narrow spectral width
($\omega\to 0$), the angular distribution $g(\theta,\theta')$ 
approaches $f(\theta,-\theta)$
and maximal entanglement in momentum is thus achieved.
The relative phase in Eq.~(\ref{eq:state}) can be written as
$$\Phi(\theta,\theta')=\Phi_{\D}(\theta,\theta')+\Phi_{\PP}(\theta,\theta')
+ \phi^{(s)}(\theta)+\phi^{(i)}(\theta')\,.$$ 
The first phase term, which 
can be expanded to the first order as
$$\Phi_{\D}(\theta,\theta')=\eta(\theta-\theta')+\Phi_0\,,$$ 
comes from the angle-dependent optical path of vertically
polarized photon pairs, generated in the first crystal, which must
travel along the second one. 
These angular dependent terms are
responsible for decoherence of polarization qubit and should be removed
in order to obtain an effective source of entangled
pairs~\cite{kwi09,cia10APL}. In our apparatus, a one dimensional
programmable spatial light modulator (SLM) is placed on the signal and
the idler paths (see Fig. \ref{exp_setup}) in order to insert the phase
functions $\Phi_{\PP}(\theta,\theta')$, $\phi^{(s)}(\theta)$, and
$\phi^{(i)}(\theta')$. The first term is used for purification, i.e to
remove the angle-dependent phase-shift, $\Phi_{\PP}=-\Phi_{\D}$, thus
realizing a reliable polarization-entanglement source~\cite{cia10APL},
whereas the functions $\phi^{(s)}$ and $\phi^{(i)}$ represent the
phase-objects we insert on the signal and the idler arm, respectively.
Notice that spatial light modulators have been recently used in imaging
also for a different purpose, i.e that of imposing a set of known random
phase distributions \cite{bro09} and implement the so-called
computational ghost imaging \cite{sha08,cle10}, where the intensity
detected in the one of the beam is computed offline.
\par
Given the state in Eq. (\ref{eq:state}), we can write the
probability for the detection of a photon pair within the emission angle
range $\theta\in(\theta_0-\frac{\delta}{2},\theta_0+\frac{\delta}{2})$,
$\theta'\in(\theta_0'-\frac{\delta'}{2},\theta_0'+\frac{\delta'}{2})$
and with polarization angles $P$ (signal) and $P'$ (idler) 
as follows
\begin{align}
C_{\hbox{\small \sc pp}^\prime}^{\delta\delta'}(\theta_0,\theta_0') 
= &\int_{\theta_0'-\frac{\delta'}{2}}^{\theta_0'+
\frac{\delta'}{2}}\!d\theta\int_{\theta_0-
\frac{\delta}{2}}^{\theta_0+\frac{\delta}{2}}\!d\theta'\,
\big\vert\langle P\theta\vert\langle P'\theta'\vert\psi\rangle\big\vert^2
\\
= &\int_{\theta_0'-\frac{\delta'}{2}}^{\theta_0'+
\frac{\delta'}{2}}\!d\theta'\int_{\theta_0-
\frac{\delta}{2}}^{\theta_0+\frac{\delta}{2}}\!d\theta\,
\left|g(\theta,\theta')\right|^2\,
\label{eq:C}
\\ &\times
\left|\cos P\cos P'
+e^{\imath \Phi(\theta,\theta')}
\sin P\sin P' 
\right|^2
\notag
\end{align}
In order to quantify the entanglement of our state we measure 
the state visibility, which is defined from Eq.~(\ref{eq:C}) 
by placing the slits on the SPDC central emission angles, 
$\theta_0,\theta_0'=0$, and setting the aperture to 
$\Delta=10\,$mrad, and using the two pairs of angles 
$P_1=P_1^\prime\equiv \alpha = \pi/4$ and $P_2=\alpha$, 
$P_2^\prime=\alpha+\pi/2\equiv\beta$, i.e
\begin{equation}
V=\frac{C_{\alpha\alpha}^{\Delta\Delta}(0,0)-C_{
\alpha\beta}^{\Delta\Delta}(0,0)}{C_{\alpha\alpha}^{
\Delta\Delta}(0,0)+C_{\alpha\beta}^{\Delta\Delta}(0,0)}\,.
\label{eq:vis}
\end{equation}
In fact, once the state has been purified and the phase objects 
$\phi^{(s,i)}$ have been inserted, the polarization density 
matrix reads 
\begin{align}
\varrho=\frac12&
\big(\ket{\HH}\bra{\HH}+\varepsilon
[\phi,\phi^\prime]\,\ket{\VV}\bra{\HH} \notag \\ &+
\varepsilon^*[\phi,\phi^\prime]\,\ket{\HH}
\bra{\VV}+\ket{\VV}\bra{\VV}\big)
\,,
\end{align}
where
$\varepsilon\equiv\varepsilon[\phi,\phi^\prime]$
is given by
$$\varepsilon=\!\int\!d\theta\,d\theta' \left|
g(\theta,\theta')\right|^2 \exp\{\imath
[\phi^{(s)}(\theta)+\phi^{(i)}(\theta')]\}\,.$$ 
Since the angular distribution $g(\theta,\theta')$ is symmetric (see
below) by choosing, without loss of generality, odd phase functions
$\phi^{(s)}$ and $\phi^{(i)}$, we obtain that 
$\varepsilon$ is real. As a consequence we may write $$\varrho=\varepsilon 
\varrho_{b}+(1- \varepsilon) \varrho_{m}\,,$$ where $\varrho_{b}$ 
denotes a Bell state and $\varrho_m$ the corresponding mixture. 
For the state
$\varrho$, visibility provides a proper measure of entanglement 
since the expression in Eq.~(\ref{eq:vis}) reduces to
$V=\hbox{Re}[\varepsilon]\equiv\varepsilon$ which, in turn, equals
the concurrence of the state.
\par
The experimental setup is shown in Fig.~\ref{exp_setup}: a linearly
polarized cw, $405\,$nm laser diode (Thorlabs LQC$405$-40P) pumps a
couple of $1\,$mm thick BBO crystals cut for Type-I downconversion. 
The beam waist is set to $\simeq500\mu$m by a telescopic
system. The effective pump power on the generating crystals is of about
$11$mW.  The $\ket{\HH}$ and$\ket{\VV}$ pairs are balanced by a half
wave-plate set along the pump path. A BBO crystal with the proper length
and optical axis angle is set on the pump path, and is used to
counteract the decoherence effect due to the temporal delay between the
two components
\cite{ser00,nam02,bar04,bri08,kwi09,cia10,cia10APL,cia08}. Such crystal
introduces a delay time between the horizontal and vertical polarization
of the pump which precompensates the delay time between the $\ket{\VV}$
pair generated by the first crystal and the $\ket{\HH}$ pair from the
second one. Signal and idler beams travel through the SLM and are
spatially selected by two irises and two slits set at $D=500\,$mm. The
low quantum efficiency of our detectors ($\sim 10\%$) forces us to
couple large angular regions: we set $\Delta x=5\,$mm which corresponds
to $\Delta=10\,$mrad.  An interference filter IF (FWHM $10\,$nm) is set
on the signal arm and selects the spectral width, whereas a long-pass
filter set on the idler arm (cut-on wavelength $715\,$nm) is used to
reduce the background. Photons are focused in two multi-mode fibers
(MMF) and sent to home-made single-photon counting modules, based on an
avalanche photodiode operated in Geiger mode with passive quenching. In
order to measure the visibility, we insert two polarizers, set at the
$P_1=P_1^\prime\equiv \alpha = \pi/4$, corresponding to a maximum in 
the coincidence rate, and $P_2=\alpha$, 
$P_2^\prime=\alpha+\pi/2\equiv\beta$ for the minimum. 
For the tomographic reconstruction (see below) 
we insert on both paths a quarter-wave plate, a half-wave plate and a 
polarizer.
\subsection{Spatial entanglement}
In our setup, the purification provided by the SLM allows us to generate 
good polarization-entangled states with visibility up to
$V=0.912\pm0.007$, which may be further increase by spatially filtering
the pump to achieve a Gaussian profile.
However, ghost imaging also require spatial entanglement and this can be
obtained upon exploiting Eq. (\ref{eq:theta-omega}), i.e. by narrowing 
the output spectral range. We use an interference 
filter, whose action is denoted by $T[\omega]$ in Eq. (\ref{eq:state}),
which selects a range of about $10\,$nm within the overall downconversion 
spectrum ($\sim200\,$nm). 
\begin{figure}[h!]
\includegraphics[width=0.92\columnwidth]{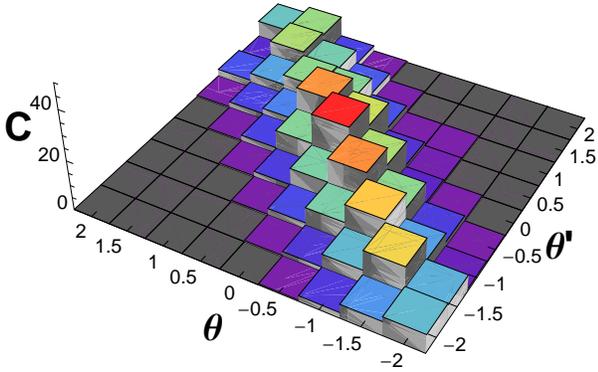}
\caption{(Color online) 
Effect of the interference filter to narrow
the output spectral range and achieve spatial entanglement:
The plot shows
coincidence counts distribution $C$ 
with $\Delta$ = $1\,$mrad and polarizers at $P=P'=H$ [see Eq.
(\ref{profHH})], measured
within a coincidence time window of $50$ns, as a function of the signal
and idler slit positions $\theta$ and $\theta'$. The phase matching
central angles correspond to $\theta_0, \theta_0'=0$.}\label{profili}
\end{figure}
\par
In order to observe the resulting angular correlation 
we place two slits of aperture $\Delta$ = $1\,$mrad, two polarizers 
at $P=P'=H$ along the downconversion arms and we measure 
\begin{align}
C &\equiv C_{\hbox{\scriptsize{HH}}}^{\Delta\Delta}(\theta_0,\theta_0')
=
\int_{\theta_0'-\frac{\Delta}{2}}^{\theta_0'+
\frac{\Delta}{2}}\!d\theta'\int_{\theta_0-
\frac{\Delta}{2}}^{\theta_0+\frac{\Delta}{2}}\!d\theta\,
\left|g(\theta,\theta')\right|^2\, 
\label{profHH}
\,,\end{align}
at $\theta_0,\theta_0' =-2\Delta,-1.5\Delta,\dots,+2\Delta$.
The phase matching (central) angles correspond to $\theta_0,
\theta_0'=0$. Coincidences are taken over an acquisition time of $6$s 
within a coincidence time window of $50$ns. 
The coincidence counts $C$ are reported in Fig. \ref{profili}
and the experimental results confirm that $g(\theta,\theta')$ is
approaching $f(\theta,-\theta)$ for the selected (narrow) spectral range.
In an analogue way we have also measured 
$C_{\hbox{\scriptsize{VV}}}^{\Delta\Delta}(\theta_0,\theta_0')$, 
checking experimentally that 
$$C_{\hbox{\scriptsize{VV}}}^{\Delta\Delta}(\theta_0,\theta_0')
\simeq C_{\hbox{\scriptsize{HH}}}^{\Delta\Delta}(\theta_0,\theta_0')\,.$$
\section{Phase imaging and nonlocal phase compensation}\label{s:dpi}
\begin{figure}[h!]
\includegraphics[width=0.99\columnwidth]{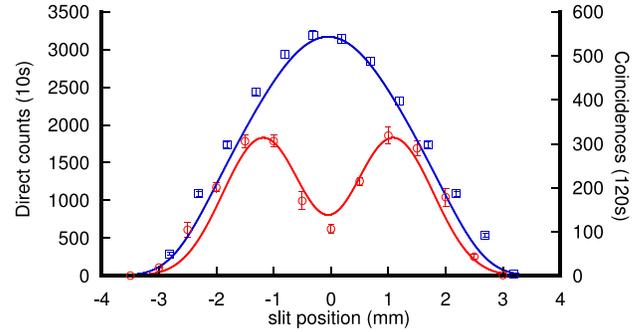}
\caption{(Color online) Phase imaging after inserting the phase 
function $\phi^{(s)}(\theta)=a \sin(k\theta)$ on the signal beam, with
$a=1.35\,$rad and $k\simeq 0.57\,$mrad$^{-1}$. Blue squares are the 
direct counts whereas red circles are the coincidences (red circles).
Solid lines are the corresponding theoretical predictions (blue and red 
lines respectively) as a function of the idler slit position. From the 
coincidence counts observed by scanning the idler beam, one can 
reconstruct the phase function along the signal beam.}\label{f:ghost-im}
\end{figure}

Single phase-object imaging consists in setting $\phi^{(i)}=0$ and 
reconstruct the phase function $\phi^{(s)}(\theta)$ inserted along 
the signal arm by scanning the coincidences for different emission 
angles $\theta'$ on the idler arm~\cite{serg04}. Experimentally, we 
insert the phase function $$\phi^{(s)}(\theta)=a \sin(k\theta)$$ using the 
SLM, with $a=1.35\,$rad and $k\simeq 0.57\,$mrad$^{-1}$. In
Fig.~\ref{f:ghost-im} we present the results for the ghost imaging:
polarizers are set to $\alpha=\pi/4$ on the signal and $\beta=3\pi/4$ 
on the idler, the slits aperture are $\Delta=10\,$mrad for the signal and
$\delta=1\,$mrad for the idler. The signal slit is centered on
$\theta_0=0$, while the idler slit varies over different values of
$\theta_0'$. The experimental data, already subtracted of the 
accidental coincidences (coincidence window
equal to $50$ns, acquisition time $120$s ) are the red circles, whereas 
the red solid line is the theoretical prediction as obtained from 
Eq.~(\ref{eq:C}), i.e.
\begin{align}
C_{\alpha\beta}^{\Delta\delta}(0,\theta_0')=\int_{\theta_0'
-\frac{\delta}{2}}^{\theta_0'+\frac{\delta}{2}}\!d\theta'
\int_{-\frac{\Delta}{2}}^{+\frac{\Delta}{2}}\!&d\theta\;\vert
g(\theta,\theta')\vert^2 \\ 
& \times \sin^2\left[\frac12\,\phi^{(s)}(\theta)\right]\notag\,.
\end{align} 
For comparison we also report the measured value of the direct counts 
(blue squares, acquisition time $10$s) with the corresponding theoretical 
prediction (solid blue line), i.e.
$$
C_{\alpha\beta}^{\Delta\delta}(0,\theta_0')=
\int_{\theta_0'-\frac{\delta}{2}}^{\theta_0'+\frac{\delta}{2}}
\!d\theta'\int_{-\frac{\Delta}{2}}^{+\frac{\Delta}{2}}\!d\theta\;\vert
g(\theta,\theta')\vert^2\,.$$ 
\par
As a matter of fact, the insertion of a single 
phase-object leads to the generation of a set of maximally entangled 
states at different angles, each one with a different phase term 
$\phi^{(s)}(\theta)$. The entanglement of the state over a broad angular 
region is thus reduced to $V=0.531\pm 0.008$. 
\par
Once we have reconstructed the signal phase $\phi^{(s)}(\theta)$ we
may further tune entanglement by imposing the phase functions 
$\phi^{(i)}=\pm\phi^{(s)}$ on the idler beam. In this way we nonlocally
superimpose two phase object. 
We remark
that, given the correlation condition $\theta'\simeq -\theta$ (see
Fig.~\ref{profili}), the overall phase function inserted by 
the SLM is given by 
\begin{align}
\phi^{(s)} (\theta)+\phi^{(i)} (\theta^\prime)&\simeq 2a\sin(k\theta) \;
\qquad \hbox{if}\quad \phi^{(i)}=-\phi^{(s)}\,,
\notag \\ 
\phi^{(s)} (\theta)+\phi^{(i)} (\theta^\prime)&\simeq 0 \qquad\qquad\qquad 
\hbox{if}\quad \phi^{(i)}=\phi^{(s)}\,. \notag
\end{align}
The corresponding visibility values are $V=0.057\pm 0.014$ 
and $V=0.888\pm 0.003$, respectively. The visibility 
of the entangled state is sligthly lower than the original 
value $V=0.912\pm0.007$ since the two beams, though showing 
high angular correlations, are not delta-correlated.
\par
In order to fully characterize the output state, and confirm that
visibility is a good figure of merit to discriminate phase functions, 
we have also performed state reconstruction by polarization qubit 
tomography for the two different output states. 
The procedure goes as follows: we measure a suitable 
set of independent two-qubit projectors~\cite{mlik00,jam01} and then 
reconstruct the density matrix from the experimental distributions 
using maximum-likelihood reconstruction. The tomographic measurements 
are obtained by inserting a quarter-wave plate, a half-wave plate and 
a polarizer. 
\begin{figure}[ht!]
\includegraphics[width=0.4\columnwidth]{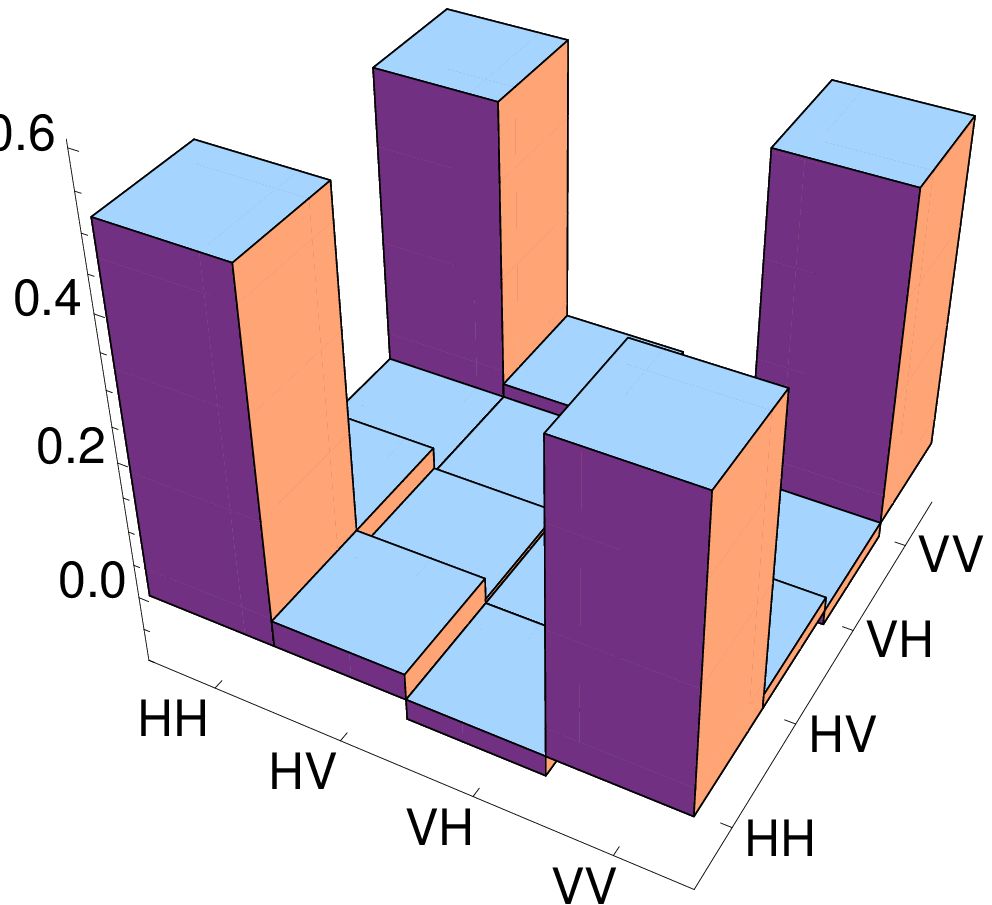}
\includegraphics[width=0.4\columnwidth]{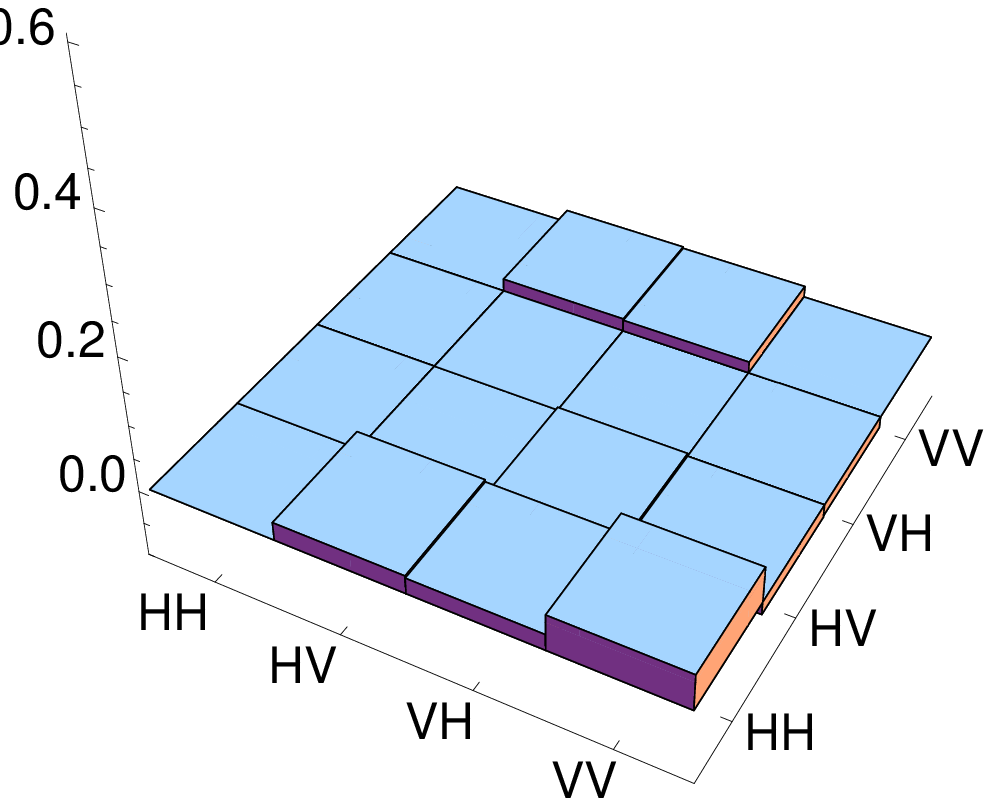}
\includegraphics[width=0.4\columnwidth]{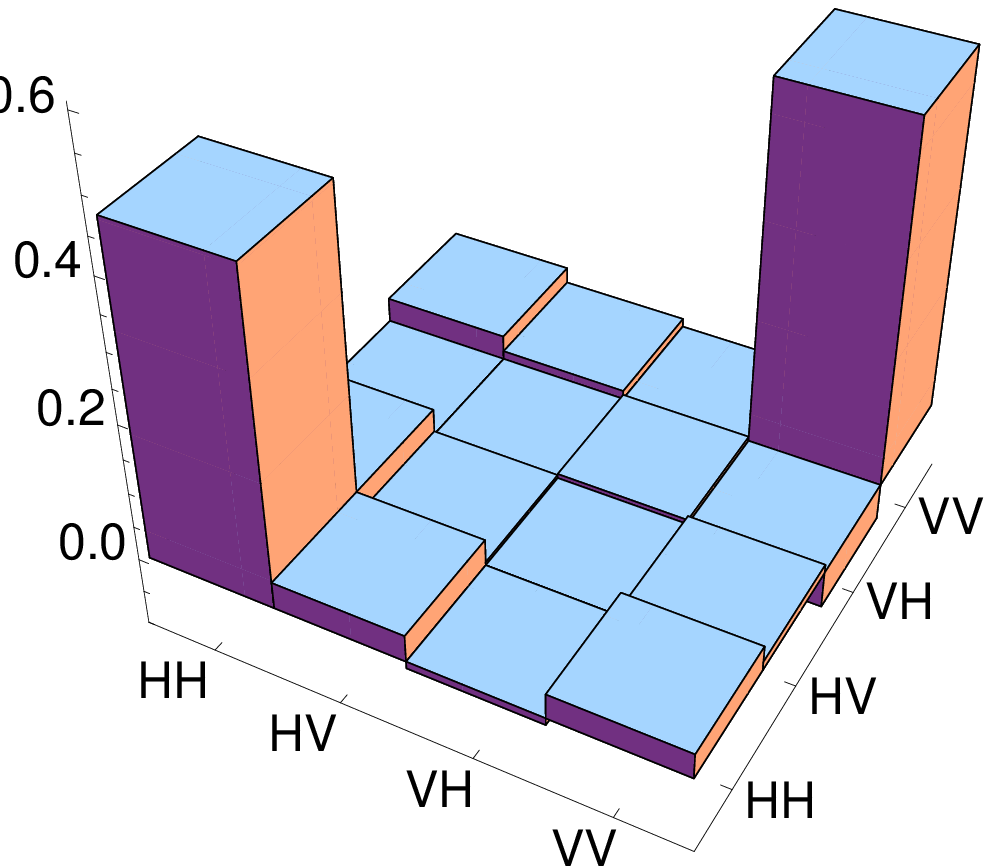}
\includegraphics[width=0.4\columnwidth]{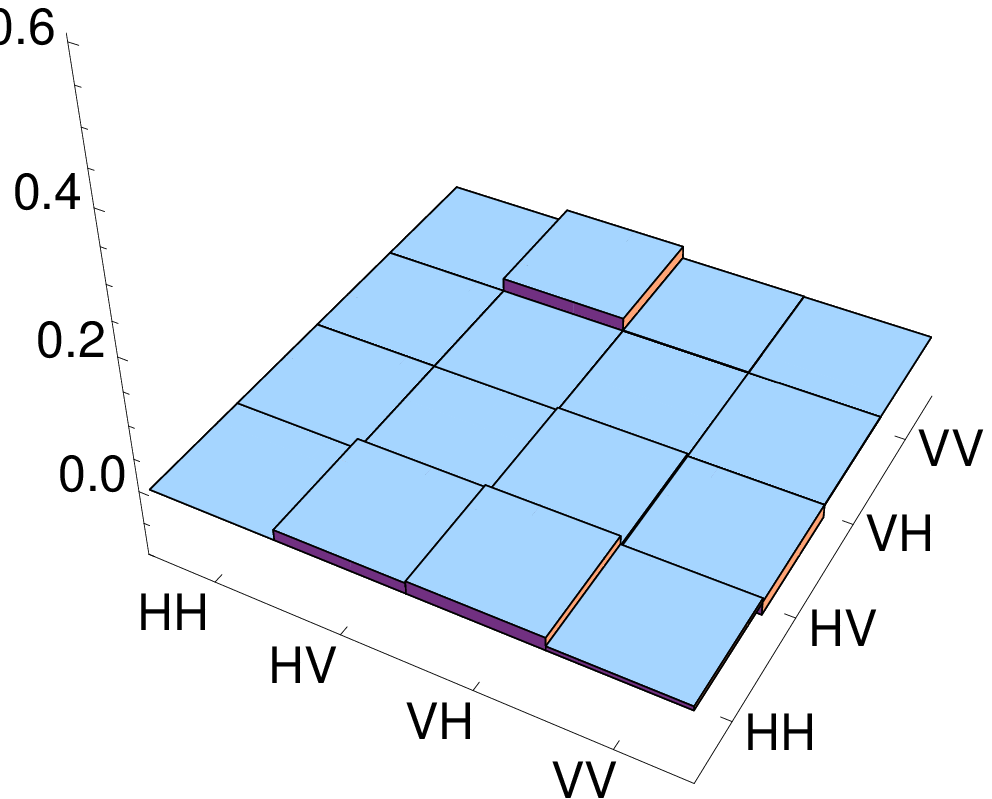}
\caption{(Color online) Tomographic reconstruction 
(the real part of the density matrix on the left 
and the imaginary one on the right) of the state in which
we insert $\phi^{(i)}=\phi^{(s)}=a \sin(k\theta)$, with $a=1.35\,$rad
and $k\simeq 0.57\,$mrad$^{-1}$ (upper panels) and of the state obtained 
with $\phi^{(i)}=-\phi^{(s)}$ (lower panels). The corresponding 
visibilities are given by $0.888\pm 0.003$ 
and $0.057 \pm 0.014$, while the measured CHSH-Bell 
parameters are $B=2.658\pm0.011$ and $B=1.854\pm
0.012$ respectively. }\label{fig:qtomo}
\end{figure}
\par
In Fig.~\ref{fig:qtomo} we present the tomographic 
reconstructions of the density matrix of the two output states: 
the reconstruction of the state resulting from the overall phase 
function $\phi^{(i)}=\phi^{(s)}$ is reported in the upper panel (real part 
on the left, and imaginary part on the right). In the lower panel we 
show the tomographic reconstruction of the density matrix for the state 
obtained with $\phi^{(i)}=-\phi^{(s)}$ (real part on the left, and imaginary part 
on the right). In order to detect the presence of 
nonlocal correlations we also measured the Bell-CHSH parameter 
$$B=\left|E(\beta_1,\beta_2) + E(\beta_1,\beta_2^\prime) + 
E(\beta_1^\prime,\beta_2)- E(\beta_1^\prime,\beta_2^\prime) 
\right|\,,$$ 
where $E(\beta_1,\beta_2)$ denotes the correlations between 
measurements performed at polarization angle $\beta_j$ for the 
mode $j$. We found 
\begin{align}
B&=1.854\pm0.012 \qquad \hbox{if}\quad \phi^{(i)}=-\phi^{(s)}\,, \notag\\
B&=2.658\pm0.011 \qquad \hbox{if}\quad \phi^{(i)}=\phi^{(s)}\,, \notag 
\end{align}
i.e.  we have violation of CHSH-Bell inequality~\cite{CHSH} by more than $57$ 
standard deviations for $\phi^{(i)}=\phi^{(s)}$, whereas no violation of 
the CHSH-Bell inequality is found for 
$\phi^{(i)}=-\phi^{(s)}$, for which the measures $B$ is less than 
the threshold $B=2$ by $12$ standard deviations.
\section{A QKD protocol based on nonlocal phase compensation}\label{s:qkd}
As a possible application of nonlocal phase compensation, we suggest to
exploit the switch between entangled and mixed states for quantum key 
distribution. Our proposal is based on the fact that
Alice and Bob control the signal and the idler arm respectively
of the downconversion output. They are thus able to insert
independently, and in a random sequence,  
the phase functions $\phi^{(A)}$ and $\phi^{(B)}$, where $\phi^{(A,B)}=\pm
a\sin(k\theta)$. In turn, this may be used to establish a 
QKD protocol as the analogue of the random choice of the signal basis or
of the measurement basis.
Alice then encodes the key ($0,1$) by adding a
constant phase ($\varphi_A=0,\pi$) to $\phi^{(A)}$. 
Upon setting the detection polarizers to $\alpha=\pi/4$ and
$\beta=3\pi/4$ we have that for $\phi^{(B)}=\phi^{(A)}$ a highly 
entangled state is shared, and thus we have a maximum in the
coincidence counting rate when $\varphi_A=\pi$ and a minimum 
when $\varphi_A=0$.  On the other hand, if $\phi^{(B)}=-\phi^{(A)}$ 
(or if an eavesdropper tries to acquire knowledge about the key), then 
Alice and Bob share a mixed (or partially mixed) state, and an intermediate
counting rate is detected. 
\begin{figure}[ht!]
\includegraphics[width=0.92\columnwidth]{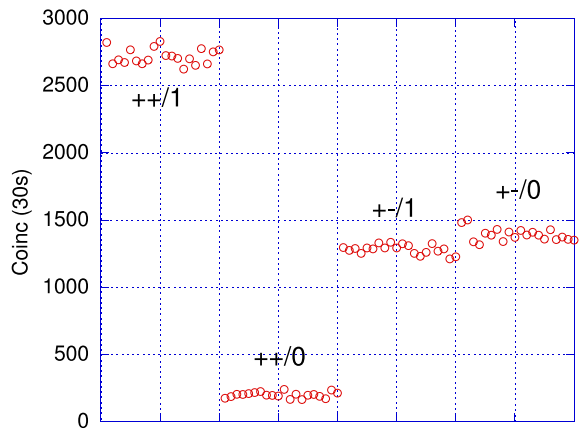}
\caption{(Color online) Quantum key distribution by nonlocal phase
compensation. The four blocks of data in the plot are the coincidence
counting rates in the four different 
configurations $[++/1]$, $[++/0]$, $[+-/1]$, and $[+-/0]$ respectively
The first two configurations correspond to the faithful transmission of
the key, whereas the two others cases are the events that Alice and Bob
are going to reject in the sifting stage of the protocol.}
\label{fig:qkd}
\end{figure}
\par
In Fig.~\ref{fig:qkd} we present the experimental results for 
the coincidence counting rate in the four possible configurations, 
which are summarized in Table \ref{tabc}, and labelled as follows: $[++/1]$ denotes
the case $\phi^{(A)}=\phi^{(B)}=a\sin(k\theta)$ and $\varphi_A=\pi$, 
corresponding to the faithful transmission of the symbol "1"; analogously
$[++/0]$ denotes the case $\phi^{(A)}=\phi^{(B)}=a\sin(k\theta)$ and $\varphi_A=0$,
leading to the faithful transmission of the symbol "0". The events that Alice and 
Bob are rejecting in the sifting stage are $[+-/0]$ and $[+-/1]$
corresponding to
$\phi^{(A)}=-\phi^{(B)}=a\sin(k\theta)$ and $\varphi_A=0,\pi$ respectively.  
\begin{table}[h!]
\caption{A QKD protocol based on nonlocal phase compensation:
the phase functions $\phi^{(A)}$ and $\phi^{(B)}$ play the role
of the random choice of the signal basis, whereas 
the key ($0,1$) is encoded by adding a
constant phase $\varphi_A=0,\pi$ to $\phi^{(A)}$. 
When the phase functions are matched, $\phi^{(A)}=\phi^{(B)}$,
we have the transmission of the key symbols.
}
\begin{tabular}{ccccc}
\hline
configuration & $\phi^{(A)}$ &$\phi^{(B)}$ &$\varphi_A$ & transmitted
key \\
\hline
$[++/1]$ & $a\sin (k\theta)$ &$a\sin (k\theta)$ & $\pi$ & "1" \\
$[++/0]$ &$a\sin (k\theta)$ & $a\sin (k\theta)$& $0$& "0"\\
$[+-/1]$ & $a\sin (k\theta)$&$-a\sin (k\theta)$ &$\pi$ & none \\
$[+-/0]$ & $a\sin (k\theta)$&$-a\sin (k\theta)$ & $0$& none \\
\hline
\end{tabular}
\label{tabc}
\end{table}
\par
Notice that the purification procedure, which removes the angular phase terms 
$\Phi_{\D}(\theta,\theta')$, allows us to generate high-quality 
entanglement states even when broad angular regions are coupled. Indeed, 
the measurements used for quantum key distribution has been performed with 
a larger aperture, in order to compensate the low quantum efficiency of 
photodetectors and to increase the measurement rate.
\section{Conclusions}\label{s:out}
In conclusion, we have suggested and demonstrated experimentally an
entanglement-based  scheme to achieve coherent nonlocal
compensation of pure phase objects. Our scheme is based on creating 
two-photon polarization and momentum entangled states where the 
insertion of a single phase object on one of the beams reduces both 
the purity of the state and the amount of shared
entanglement, and where the original entanglement can be retrieved by
adding a suitable phase object on the other beam.  
In our setup polarization and momentum entangled
states are generated by spontaneous parametric downconversion and
then purified using a programmable spatial light modulator. The same
device is also used to impose arbitrary space dependent phase 
functions to the beams, which play the role of arbitrary pure phase
objects. Finally, we have suggested a novel quantum key distribution 
protocol exploiting the effect of nonlocal phase compensation and 
we have provided its experimental verification.
Our results prove experimentally the feasiblity of 
coherent nonlocal compensation/superposition of pure phase 
objects and pave the way for further developments, as the 
reconstruction of the overall phase function imposed 
to the two beams.
\section{Acknowledgements}
MGAP thanks Alessandra Gatti and Konrad Banaszek for useful discussions.

\end{document}